\documentstyle[preprint,prb,aps]{revtex}

\begin{document}
\draft
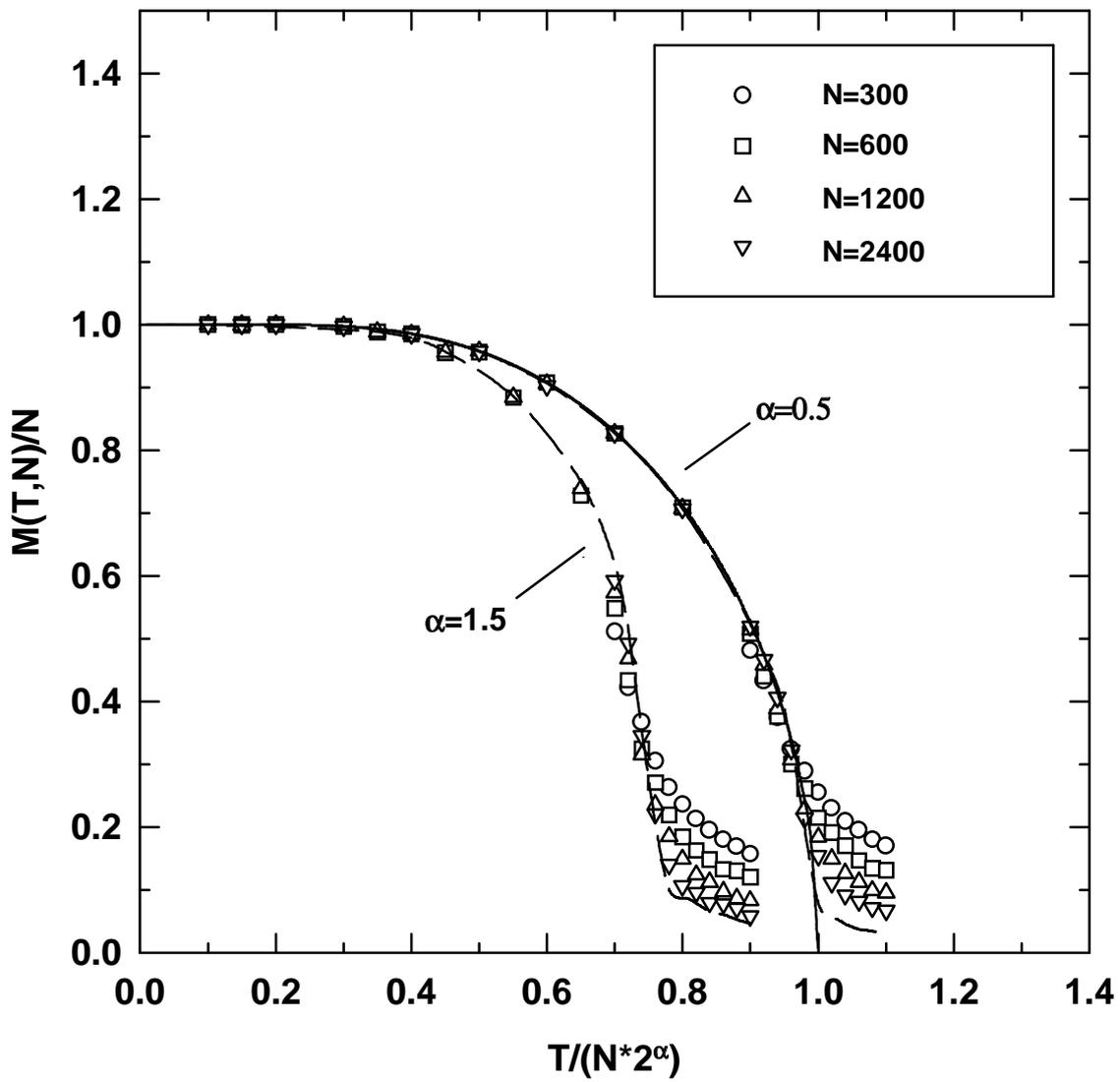
\title{Long-range interactions and non-extensivity
in ferromagnetic spin models}

\author{Sergio A. Cannas\cite{auth} and Francisco A. Tamarit}  

\address{Facultad de Matem\'atica, Astronom\'\i a y F\'\i sica, 
Universidad Nacional de C\'ordoba, Haya de la Torre y Medina Allende
S/N, Ciudad Universitaria, 5000 C\'ordoba, Argentina
\cite{auth2}\cite{auth3} }

\date{\today}
\maketitle

\begin{abstract} 
The Ising model with ferromagnetic interactions that decay as
$1/r^\alpha$ is analyzed in the non-extensive regime $0\leq\alpha\leq
d$, where the thermodynamic limit is not defined.
In order to study the asymptotic properties of the model in the
$N\rightarrow\infty$ limit ($N$ being the number of spins) we propose
a generalization of the Curie-Weiss model, for which the
$N\rightarrow\infty$ limit is well defined for all $\alpha\geq 0$. We
conjecture that mean field theory is {\it exact} in the last model
for all $0\leq\alpha\leq d$. This conjecture is supported by Monte
Carlo heat bath simulations in the $d=1$ case. Moreover, we
confirm a recently conjectured scaling (Tsallis\cite{Tsallis}) 
which allows for a unification of extensive ($\alpha>d$) and
non-extensive ($0\leq\alpha\leq d$) regimes.
\end{abstract}


It has been known for a long time that systems with long-range
microscopic interactions can exhibit non-extensive behaviour
(see Refs.[1-3], among many others, and references therein). In
other words, if the effective range of the interactions between
the constituent particles decays slowly enough with the
distance, the free energy $F = -\beta \ln{Z}$, with $Z\equiv
Tr\, exp(-\beta\,H)$ ($H$ being the Hamiltonian of the system
and $\beta\equiv k_BT$) will grow {\it faster} than the number
$N$ of microscopic elements when $N\rightarrow \infty$ and the
so-called thermodynamic limit will be not defined.

Besides their fundamental theoretical interest in physics,
microscopic models with long-range interactions which decay
slowly are of interest nowadays, in view of their relationship
with neural systems modeling \cite{Amit}, where far away
localized neurons interact through an action potential that
decays slowly along the axon. Another related problems,
are spin sytems with RKKY like interactions, which are present
in spin glasses \cite{Ford}, critical phenomena in highly ionic
systems \cite{Pitzer}, Casimir forces in fluid near the critical
point \cite{Burkhardt} and phase segregation in model alloys
\cite{Giacomini}. Many of these problems can be studied using some
variation of the Ising model (e.g. Hopfield model of neural network,
Edward-Anderson of spin glasses, etc.), or its lattice--gas version,
as in model alloys \cite{Giacomini}. Moreover, even systems not
directly related with magnetic ones, present often critical
properties that fall in the universality class of some magnetic
systems, the Ising model being the most simple non--trivial one.
Hence, a deep comprehension of the general properties of the
Ising model with long--range interactions is relevant to understand
the behaviour of this kind of systems. As we will show, even the most
simple case, {\it i.e.} the ferromagnetic model, presents non-trivial
non-extensive behaviours and therefore it represents a good starting
point to the study of more complex models. 

In this letter we consider an Ising ferrromagnet with {\it long-range}
interactions, that means, a system described by the Hamiltonian

\begin{equation}
H = -\sum_{(i,j)} J(r_{ij})\; S_i S_j 
\,\,\,\,\,\,\,\,\, \text{($S_i=\pm 1$\,\, $\forall$ $i$)}
\label{H1}
\end{equation}

with 

\begin{equation}
J(r_{ij})=\frac{J}{r_{ij}^\alpha}
\,\,\,\,\,\,\,\,\, \text{($J>0$; $\alpha >0$)}
\end{equation}

\noindent where $r_{ij}$ is the distance (in crystal units) between
sites $i$ and $j$, and where the sum $\sum_{(i,j)}$ runs over all
distinct pairs of sites on a $d$-dimensional simple hypercubic
lattice.  The $\alpha \rightarrow \infty$ limit corresponds to the
first-neighbor model. The $\alpha=0$ limit corresponds, after a
rescaling $J\rightarrow J/N$, to the Curie-Weiss model.

Let us introduce the sums $\phi_i(\alpha)=\sum_{j\neq i} J(r_{ij})$.
A sufficient condition (and believed to be necessary\cite{Thompson})
for the existence of the thermodynamic limit of this system is that

\begin{equation}
\phi(\alpha)= \lim_{N\rightarrow\infty} \frac{1}{N} \sum_i
\phi_i(\alpha) < \infty.
\label{phi}
\end{equation}

Let us now take a d-dimensional hypercube of side $L+1$ and
$N=(L+1)^d$, and let $i=0$ be the central site of the hypercube. We
have that 
\begin{equation}
\phi(\alpha)=\lim_{N\rightarrow\infty} \phi_0(\alpha).
\label{phi0}
\end{equation}

\noindent Then 

\begin{equation}
\phi_0(\alpha)= J\, \sum_{i_1=1}^{L/2} \cdots
\sum_{i_d=1}^{L/2} \frac{1}{(i_1^2+i_2^2+\cdots+i_d^2)^{\alpha/2}}.
\end{equation}

Using Euler-McLaurin sum formula \cite{Bruijn} we can approximate, 
for $L \gg 1$,

\begin{eqnarray}
\phi_0(\alpha)&\approx& J\, 2^d\, \int_{1}^{L/2} dx_1 \;\cdots
\int_{1}^{L/2} dx_d\;\frac{1}{(x_1^2+x_2^2+\cdots+x_d^2)^{\alpha/2}}
\nonumber \\
\mbox{ }      &\propto& J\,  \int_{1}^{L/2} dr\; r^{d-1-\alpha}.
\nonumber
\end{eqnarray}

Hence, $\phi_{0}(\alpha)$ shows the following asymptotic behaviour
for $N \gg 1$:

\begin{equation}
\phi_0(\alpha) \sim J\, C_d(\alpha) 2^\alpha \left\{
\begin{array}{ll}
\frac{1}{1-\alpha/d}\left(N^{1-\alpha/d}-1\right)& \;\;\;\text{if
$\alpha\neq d$} \\
\ln{N} & \;\;\;\text{if $\alpha = d$}
\end{array} \right.
\label{phi2}
\end{equation}.

In other words:

\begin{equation}
\lim_{N\rightarrow\infty} \frac{(1-\alpha
/d)}{\left(N^{1-\alpha/d}-1\right) J 2^{\alpha} } \phi_{0}(\alpha) =
C_{d}(\alpha) \;\;\;\; \text{for}\;\;\;\;\ \alpha \neq d
\end{equation}

\noindent and

\begin{equation}
\lim_{N\rightarrow\infty} \frac{ \phi_{0}(d)}{J 2^{d} \ln{N}} = 
C_{d}(d) 
\end{equation}

\noindent where $C_d(\alpha)$ is a continuous function of $\alpha$
independent of $N$, with $C_d(0)=1$. Therefore, the thermodynamic
limit is well defined for $\alpha>d$ (where the system presents
extensive behaviour), while for $\alpha \leq d$ the system becomes
non-extensive, the critical temperature becomes infinite and the
standard Maxwell-Boltzmann formalism cannot be
applied\cite{Hiley,Ainzenman}.  The system undergoes a second order
phase transition at finite temperature for all $\alpha>d$
when\cite{Hiley}  $d\geq 2$ and for $1\leq\alpha\leq 2$
when\cite{Cannas} $d=1$.  For $\alpha
\rightarrow d^+$, the critical temperature shows the following
asymptotic behaviour \cite{Hiley}:

\begin{equation}
k_BT_c \sim J\, \phi(\alpha)
\end{equation}

We now introduce a new model that generalizes the Curie-Weiss one.
Such model is described by the Hamiltonian:

\begin{equation}
H' = - \sum_{(i,j)} J'(r_{ij})\; S_i S_j 
\label{H2}
\end{equation}

\noindent with 

\begin{equation}
J'(r_{ij})=\frac{J(r_{ij})}{N^*(\alpha) 2^\alpha}
\end{equation}

\noindent where

\begin{equation}
N^*(\alpha) = \frac{1}{1-\alpha/d} (N^{1-\alpha/d}-1)
\end{equation}

\noindent which behaves as 

\begin{equation}
N^*(\alpha) \sim \left\{
\begin{array}{ll}
\frac{1}{\alpha/d-1}& \;\;\;\text{for $\alpha/d > 1$} \nonumber\\
\ln{N} & \;\;\;\text{for $\alpha/d = 1$}\\
\frac{1}{1-\alpha/d} N^{1-\alpha/d}& \;\;\;
\text{for $0\leq\alpha/d\leq 1$} 
\end{array} \right.
\end{equation}

\noindent for $N\rightarrow\infty$. This model reduces to the
Curie-Weiss one for $\alpha=0$ and to our original model
Eq.(\ref{H1}) (after rescaling $J\rightarrow
J(\alpha/d-1)/2^\alpha$) for $\alpha>d$. From Eqs.(\ref{phi}),
(\ref{phi0}) and (\ref{phi2}) we see that the thermodynamic
limit of this model is well defined {\it for all} $\alpha \geq
0$.  We expect this system to show a phase transition at finite
temperature for all $\alpha\geq 0$ when $d\geq 1$ and for
$0\geq\alpha\geq 2$ when $d=1$.

The mean field theory for this model predicts a critical temperature
$k_BT'_c = J\, C_d(\alpha) 2^{\alpha}$, which is {\bf exact} for
$\alpha=0$ and for $\alpha\rightarrow d^+$. Hence, {\bf we conjecture
the critical temperature reproduces exactly the mean  field
prediction for all} $0\leq \alpha\leq d$. This conjecture is
difficult to verify for $d > 1$, since for systems with long--range
interactions it is hard to obtain reliable numerical data for the
exact critical temperature. In what follows we show that
$C_1(\alpha)=1$ for $0\leq\alpha\leq 1$ and then we will test our
conjecture through a Monte Carlo numerical simulation.

Let us consider the $d=1$ system. In this case $\phi_0=2J\,
\sum_{n=1}^{L/2} \frac{1}{n^\alpha}$.
Then, for $\alpha>1$ $\phi(\alpha)=2J\, \zeta(\alpha)$ ($\zeta(x)$ is
the Riemann Zeta function) and the critical temperature diverges
as\cite{Cannas} $k_BT_c\sim 2J/(\alpha-1)$ for
$\alpha\rightarrow 1^+$ .  Using the asymptotic behaviours
\cite{Bruijn}

\begin{equation}
\sum_{n=1}^M n^{-z} \sim \frac{M^{1-z}}{1-z}
\;\;\;\;\;\; \text{for $Re(z)>-1$ and $z\neq 1$} 
\label{asimp}
\end{equation}

\begin{equation}
\sum_{n=1}^M \frac{1}{n} \sim \ln{M}
\label{asimp2}
\end{equation}

for $M\rightarrow\infty$, we get for $\alpha<1$

\begin{equation}
\phi_0 \sim \left\{ \begin{array}{ll}
                   2^\alpha \frac{N^{1-\alpha}}{1-\alpha}&
		   \;\;\;\;\;\; \text{for $0\leq\alpha\leq 1$} \\
		   2\, \ln{N}&
		   \;\;\;\;\;\; \text{for $\alpha= 1$} \\
		   \end{array}\right.
\end{equation}

and

\begin{equation}
C_1(\alpha) = \left\{ \begin{array}{ll}
		       1&\text{for $0\leq\alpha\leq 1$} \\
		       \frac{\alpha-1}{2^{\alpha-1}}\zeta(\alpha)&
		       \text{for $\alpha>1$}
		      \end{array} \right.
\end{equation}

 For $d=1$ the following necessary condition must be satisfied in
order to have a finite critical temperature \cite{Ellis}:

\begin{equation}
\lim_{N\rightarrow\infty} \sum_{n=1}^N n\, J(n) = \infty.
\end{equation}

\noindent Using Eq.(\ref{asimp}), we see that 

\[
\sum_{n=1}^N n\, J'(n) \sim \frac{J}{N^*(\alpha)\; 2^\alpha}
\frac{N^{2-\alpha}}{2-\alpha}
\]   

\noindent Hence, the critical temperature for $d=1$ will be finite
$\forall$ $0\leq\alpha\leq 2$.

If we denote by $u'$, $s'$ and $f'$ the energy, entropy and free
energy per particle associated with the Hamiltonian $H'$, {\it i.e.},

\[ 
f'(T) = \lim_{N\rightarrow\infty} -\frac{\beta}{N} \ln{Z'},
\]

with $Z\equiv Tr_{\left\{S_i\right\}}\, exp(-\beta\,H')$,
 
\[ 
u'(T) = \lim_{N\rightarrow\infty} \frac{1}{N} Tr_{\left\{S_i\right\}}\,
H'\, exp(-\beta\,H')
\]
 
and

\[ f'(T)=u'(T)-T\, s'(T) \]

\noindent we see that the generalized thermodynamic behaviour
associated with the Hamiltonian (\ref{H1}) can be accommodated, {\it
for all} $\alpha\geq 0$, with the following scalings (in the limit
$N\rightarrow\infty$ and for $T>0$):

\begin{eqnarray}
U(N,T) &\sim& N\, N^*\, u'(T^*) \label{eqU}\\
S(N,T) &\sim& N\, s'(T^*)\\
F(N,T) &\sim& N\, N^*\, f'(T^*) \label{eqF}
\end{eqnarray}

\noindent with $T^*\equiv T/N^*$, as was recently conjectured by
Tsallis for general systems with long-range interactions
\cite{Tsallis}.  Moreover, it can be easily shown that this type of
scaling preserve the Legendre transformation structure of the
thermodynamics, even in the long-range regime\cite{Tsallis}
$0\leq\alpha\leq d$.  It is also expected that the magnetization
$M\equiv \left< \sum_i S_i \right>$ scales as $M(N,T) \sim N m(T^*)$.
Therefore, the suitable plot for looking for data collapse in a
numerical simulation will be $M(N,T)/N$ {\it vs} $T/N^*$.

Let us consider the $d=1$ case. We performed a Monte Carlo simulation
on a chain of $N$ spins with Hamiltonian
(\ref{H1}), using heat bath dynamics, for $N=75$, $150$, $300$,
$600$, $1200$ and $2400$. We calculated the root-mean-square of the
magnetization of the system $M(N,T)$ as a function of the temperature
$T$ for$\alpha=$ $0$, $0.25$, $0.5$, $0.75$ and $1.5$.  The results
were averaged 
over $K$ samples with different random number sequences ( $K=100$,
$50$, $20$, $20$, $10$ and $5$ for $N=75$, $150$, $300$, $600$,
$1200$ and $2400$ respectively).  For every value of $\alpha$ we
obtained an extrapolated magnetization curve $M_\alpha (T)$, by
performing a numerical extrapolation of $M(N,T)$ in $1/N$ to
$N\rightarrow\infty$. 

In Fig.(\ref{fig1}) we show our results for $M(N,T)/N$ {\it vs}
$T/(N^*\;2^\alpha)$ for $\alpha=0.5$ and $1.5$. These 
curves show clearly the data collapse above mentioned. Moreover, for
$0\leq\alpha<1$ our results show that all curves $M_\alpha (T)$ {\bf
fall into a single one}, which coincides with the well known exact
solution for the Curie-Weiss model ($\alpha=0$ case), {\it i.e.} the
solution of the equation $m=\tanh{(m/T')}$ ($m\equiv M/N$; $T'\equiv
T/(N^*\;2^\alpha)$. This last result is impressive. It does not only
confirm our conjecture concerning the critical temperature
($T'_c=1$), but also shows that the full equation of state $m=m(T)$
at zero magnetic field becomes independent of $\alpha$ in the
non-extensive regime $0\leq\alpha\leq 1$, suggesting that all the
thermodynamic functions are those predicted by the mean field theory.
These results are consistent with recent Monte Carlo simulations of
the correlation function, which reproduce the mean field behaviour in
the same region\cite{Bergersen} $0\leq\alpha\leq 1$.

In this letter we have found a new scaling for the Ising model with
long range interactions that allows us to get a well defined 
thermodynamic limit for any value of $\alpha$. In particular, for
$\alpha = 0$, we recover the well--known Curie-Weiss scaling, which has
been vastly used in the context of magnetic systems. With this 
scaling we were able to obtain the generalized thermodynamic behaviour
for $N \rightarrow \infty$ (Eqs. \ref{eqU}--\ref{eqF}) in the 
ferromagnetic case, which had been previously conjectured in a more
general context by C. Tsallis \cite{Tsallis}. It is worth
stressing that with this scaling both extensive and
non-extensive behaviours can be accommodated in a unified and
elegant formalism and  the (until now almost unexplored) $0 \leq \alpha
\leq d$ case becomes tractable. In the same way the non-extensive
fully connected Ising model showed to be a very useful tool when
suitable rescaled (Curie Weiss model), we believe that the model 
here analyzed can  represent not only a useful approach but also
a more realistic one for certain problems such as neural networks and
spin glasses among others. 

On the other hand, due to the distance dependence of the interactions
it becomes very difficult to obtain exact analytical results even in
the $d=1$ case. We calculated the critical temperature in the mean
field approximation for any value of $d$ and presented some numerical
evidence that, (at least for $d=1$) not only it reproduces the exact
value in the whole non-extensive regime $0 \leq \alpha \leq d=1$, but
also the full magnetization curve $M(T)/N$ is given by the mean field
one. We believe that the critical temperature satisfies that property
for all $d$. This conjecture is partially supported by the fact that
it holds for $\alpha = 0$ and $\alpha
\rightarrow d^+ $.  Moreover, since the critical exponent are those
of the mean field theory both for $\alpha = 0$ and $\alpha
\rightarrow d^+ $, we conjecture that all the critical properties
will reproduce the mean field behaviour for $0 \leq \alpha \leq d$.
Monte Carlo simulation of the correlation function \cite{Bergersen}
for $d=1$ also support this statement. These results, although
intuitive, are non-trivial and important, specially concerning spin
glasses and biological systems (neural networks, immunology, etc)
where a common approximation consists in to consider fully connected
models instead of the more realistic ones with slow decaying
interactions (e.g. RKKY). Our results show that mean field behaviour
is robust again variations of  the range of the interactions $\alpha$
whithin the non-extensive region, at least for $d=1$.  If our
conjecture were true, this would have important practical
implications: if you are considering systems with slow enough
decaying interactions then you do not need sophisticated
approximations, at least as far as critical properties are concerned!

It would be very interesting to extend the present analysis to more
general systems of interacting particles with long range interactions.

We are imdebted to  Constantino Tsallis for many fruitful suggestions
and discussions about this beautiful problem. Useful discussions with
G. Raggio and many valuable referee's suggestions are acknowledged.
We also acknowledge warm hospitality received at the Centro
Brasileiro de Pesquisas F\'\i sicas (Brazil), where this work was
partly carried out.  This work was partially supported by grants
B11487/4B009 from Funda\c{c}\~ao Vitae (Brazil), PID 3592 from
Consejo Nacional de Investigaciones Cient\'\i ficas y T\'ecnicas
CONICET (Argentina),PID 2844/93 from Consejo Provincial de
Investigaciones Cient\'\i ficas y Tecnol\'ogicas (C\'ordoba,
Argentina) and a grant from Secretar\'\i a de Ciencia y Tecnolog\'\i
a de la Universidad Nacional de C\'ordoba (Argentina).

\begin{figure}
\caption{Monte Carlo calculations of root-mean-square of the
magnetization per spin $M(N,T)/N$ {\it vs} the scaled temperature
$T/(N^*\; 2^\alpha)$ using Hamiltonian (\protect{\ref{H1}}), for
$\alpha=0.5$, $1.5$ and different values of the number of spins $N$
in the one-dimensional lattice. In all cases the error bars are
smaller than $0.01$. The dashed lines are the $N=\infty$
extrapolation of $M(N,T)/N$. The extrapolated curves for $\alpha=0$,
$0.25$ and $0.75$ are indistinguishable from the previous one. The
solid line is the exact solution of the Curie-Weiss model.}
\label{fig1}
\end{figure}

\end{document}